# Enhancing Music Genre Classification through Multi-Algorithm Analysis and User-Friendly Visualization

**Author1:** Navin Kamuni
**Email:** navin.kamuni@gmail.com
**Affiliation:** *AI ML M.Tech, BITS Pilani WILP,* USA

**Author2:** Dheerendra Panwar
**Email:** dheerendra.panwar@ieee.org
**Affiliation:** Independent Researcher, CA, USA

**Abstract**

The aim of this study is to teach an algorithm how to recognize different types of music. Users will submit songs for analysis. Since the algorithm hasn't heard these songs before, it needs to figure out what makes each song unique. It does this by breaking down the songs into different parts and studying things like rhythm, melody, and tone via supervised learning because the program learns from examples that are already labelled. One important thing to consider when classifying music is its genre, which can be quite complex. To ensure accuracy, we use five different algorithms, each working independently, to analyze the songs. This helps us get a more complete understanding of each song's characteristics. Therefore, our goal is to correctly identify the genre of each submitted song. Once the analysis is done, the results are presented using a graphing tool, making it easy for users to understand and provide feedback.

**Keywords:** Algorithm, Music, Song, Supervised Learning

## 1. Introduction

The challenges faced by music enthusiasts serve as the motivation for a pioneering research venture. The initial inspiration for this study came from encountering videos showcasing Artificial Intelligence (AI)-generated music. By exposing the model to a rich and varied corpus of musical examples, it was able to discern subtle patterns and nuances inherent to different genres, thereby honing its classification capabilities. However, upon closer examination, it became evident that existing approaches to AI-generated music were limited in their ability to produce coherent compositions without human intervention. Recently一to optimize music classification via the utilization of AI and Machine Learning (ML) has been extensively studied [1]–[3]. Once the ML model had been trained and validated, attention turned towards the development of a user-friendly interface that would facilitate interaction with the classification system. However, the goal was ambitious yet attainable一to train a ML algorithm to recognize and categorize songs with a degree of accuracy rivalling that of seasoned music enthusiasts. Central to the study's success was the incorporation of supervised learning techniques, wherein the ML model was trained on a diverse dataset comprising a wide range of musical genres and styles. With this newfound perspective, the focus shifted towards developing a ML model capable of accurately classifying music based on key features such as tempo, rhythm, key, and time signature. The culmination of these efforts resulted in the creation of a robust music classification system capable of accurately identifying the genre of a given song with a high degree of confidence. However, it became increasingly apparent that the success of the classification algorithm relied not only on the sophistication of the underlying ML techniques but also on the quality and diversity of the training data. The realization dawned that the true potential of AI in the realm of music lay not in composition but in classification. In the realm of music appreciation, enthusiasts often find themselves challenged when attempting to articulate the essence of the music they love. Therefore, the implications of this study extend far beyond the realm of music classification. TensorFlow[1], a powerful open-source ML framework, served as the cornerstone一providing a robust platform for model development and training. Thus, efforts were directed towards extracting quantitative features such as Beats Per Minute (BPM), spectral characteristics, and rhythmic patterns, which could serve as inputs to the ML model. However, this challenge is exacerbated by the proliferation of experimental music, where artists continuously push the boundaries of convention and defy traditional genre norms. One of the primary challenges encountered during the study's development was the selection and extraction of relevant features from audio files.

---

[1] https://www.tensorflow.org/

This paper is as follows in the next section we will see the related works. In Section 3, the materials and methods are discussed. In Section 4, the implementation of the framework is presented. In Section 5, the experimental analysis is conducted along with results. In Section 6, the thoughts of the paper are discussed and we conclude the paper in Section 7 with some conclusion and future works.

## 2. Related Works

In recent studies, researchers have explored various methodologies for audio classification, aiming to enhance accuracy and efficiency in this domain. Recent research efforts underscore the importance of innovative methodologies, diverse datasets, and rigorous comparative analyses in advancing the field of audio classification. Interestingly, the Convolutional Neural Network (CNN) model exhibited superior accuracy in genre classification tasks. [4] have explored diverse datasets, such as 10-second audio clips sourced from a massive compilation of 2.1 million YouTube videos. In a separate experiment, a Residual Neural Network (RNN) was employed to process 3-second audio clips extracted from the GTZAN dataset. Notably, a CNN yielded the highest accuracy of 88.5% when trained on Mel-Frequency Cepstral Coefficients (MFCC) and additional song features according to [5]. After that, [6]─experimented with different visualization techniques for audio classification, including spectrograms, chromagrams, and MFCC visualizations. These diverse approaches offer valuable insights into the underlying characteristics of audio signals. A comprehensive comparison of ML algorithms was conducted, focusing on music genre classification tasks by [7]. These datasets provided ample material for training and testing various classification models. [8] investigation delved into the comparison between CNN and Long Short-Term Memory (LSTM) models, utilizing MFCC features. In [9], a novel classifier was introduced, achieving a noteworthy accuracy rate of 65%, alongside an AUC of 0.894. The classifier leveraged the mean and co-variance of MFCC during the training process. Here is a summary of key findings from recent research endeavors. The model architecture included duplicate convolution layers followed by distinct pooling layers, alongside rigorous statistical analysis.

## 3. Materials and Methods

The objectives outlined within the study underscore the development of the application initially on Linux, with potential adaptability for mobile platforms in the future. Moreover, the application must exhibit fault tolerance, accommodating incomplete user inputs without compromising accuracy. While, optimizing the application for seamless operation on mobile platforms entails addressing compatibility and performance concerns inherent to such environments. These requirements collectively aim to enhance usability and ensure seamless interaction with the application. The aspiration is to develop an application that allows users to record musical pieces, subsequently facilitating accurate feature identification. Acknowledging the limitations of mobile platforms regarding audio classification─the study envisions the offloading of such tasks to servers, contemplating the utilization of AWS due to its robust infrastructure and comprehensive documentation. Integral to this endeavor is the seamless processing of mp3 files, facilitating accessibility and usability. To mitigate potential user frustration during data processing, ongoing feedback mechanisms, including aesthetic loading indicators, are deemed indispensable. Additionally, ensuring consistent accuracy across diverse musical inputs poses a formidable challenge, necessitating meticulous model refinement and validation. These encompass intuitive navigation within the application interface, a dedicated button to initiate the analyzation process, submission and analysis of audio files, storage of results for iterative learning, visualization of outcomes through percentage charts, provision for recording additional audio, and user-driven result labeling. Concurrently, non-functional requirements underscore ancillary aspects critical to the application's success. At the core of the discussion lies the problem statement─enabling individuals to identify musical features effortlessly, akin to human capability but with the precision and efficiency afforded by ML algorithms. Efficiency emerges as a key consideration, mandating swift user-server interactions to preempt user disengagement. The overarching goal remains high accuracy in feature identification, whether it pertains to testing data or user-recorded audio, underscoring the application's efficacy and reliability. Foremost among these is the possibility of encountering API-related issues with AWS integration.

## 4. Implementation

The implementation approach for the study involves developing a user-controlled music classification application using ML algorithms and graphical representations. The approach encompasses developing a user-controlled music classification application with a phased development approach and emphasis on risk assessment, methodology, and evaluation criteria. The core functionality of the application revolves around processing user-submitted audio through various ML algorithms to accurately classify the music into its respective genre. This application aims to utilize ML algorithms to analyze the audio and accurately determine the genre it belongs to. The development of an application─allows users to submit music and have it classified into different genres.

The architecture of the solution involves several components, including Python for algorithm scripting, Matplotlib[2] for creating graphical representations of data, and Kivy[3] for developing the application framework. Risks such as issues with audio recording and processing, displaying graphical data, processing time/cost, application bugs, and integration with cloud services, etc. ML algorithms, including k-nearest neighbor [10]–[12] and naive bayes [13]–[15], are implemented, and tested for classification accuracy. The study adopts a phased development approach, with each phase focusing on specific aspects of the application development process. For example, the initial phase involves building the framework for the application and adding basic functionality to analyze user audio. The application aims to achieve this by extracting relevant features from the audio data, such as tempo, zero crossing rate, and chromagram, and using these features as input for the classification algorithms. The prototyping phase of the study involves creating wireframes inspired by existing applications such as Shazam[4], which feature prominent recording buttons and loading screens to indicate processing. The core functionality revolves around analyzing user-submitted audio through various algorithms to classify music accurately. The study also explores the use of other classification algorithms, such as decision trees [16]–[18], random forest [19]–[21], and neural networks [22], to determine their effectiveness in classifying audio. The study aims to evaluate its success based on the core functionality of accurately classifying audio, even if certain features need to be sacrificed to achieve this goal. For example, to address the risk of long processing times due to audio recording, the study proposes limiting or specifying a fixed recording time to avoid vague answers and reduce processing time. Evaluation criteria are established to assess the success of the study, with a focus on achieving a Minimal Viable Product (MVP) that accurately classifies audio into different genres.

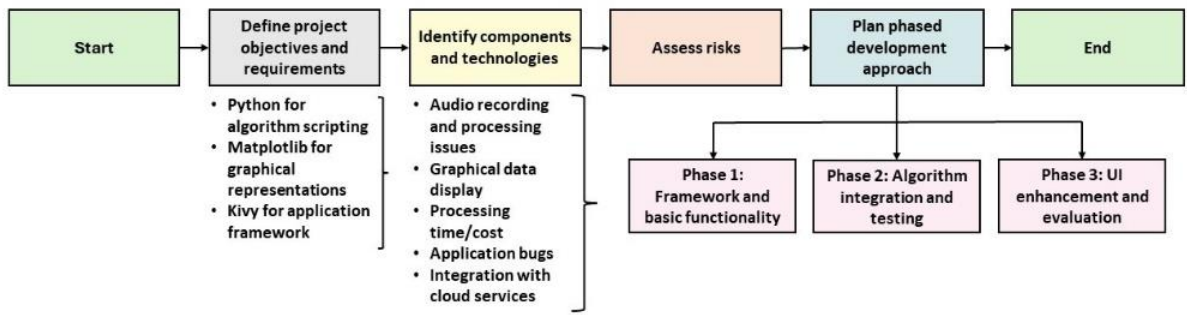

Fig. 1. Flowchart of the implementation

**4.1 Further Improvements**

Initially envisioned as a portable mobile application for music classification, the study's architecture had to be adjusted due to limitations in cloud service integration. These included aspects such as selecting the appropriate ML algorithms to ensure high accuracy, finding suitable datasets for music analysis, developing the Graphical User Interface (GUI) using Kivy, and setting up a database for accessing music information via Google API. The implementation phase of the study involved various challenges and compromises that were made to ensure the completion and functionality of the study. Consequently, the developing of a desktop Linux application that could analyze music files locally. Porting the Kivy app to Android encountered significant obstacles due to unsupported libraries, ultimately leading to the abandonment of the Android application. However, certain requirements, such as sending recordings to a server for analysis and storing user history, were not implemented due to time constraints and technical difficulties. Easy difficulties included aspects such as planning and schedule management, which were generally manageable except during periods of heavy workload. Risk assessment identified the failure of the mobile application as a significant risk, leading to adjustments in study priorities and methodologies. Hard difficulties, however, proved to be insurmountable challenges. A prototype of the application was developed using Kivy GUI (refer to Fig. 2), offering users a streamlined experience for music classification with automated feature extraction and genre classification. These difficulties were categorized into three levels—easy, medium, and hard. Agile development methodologies, such as sprint planning and backlog management, were employed to ensure progress and task completion within manageable timeframes. Medium difficulties required more effort but were ultimately resolved. Similarly, integrating cloud solutions such as AWS and Google Cloud proved overly complex, exceeding the study's scope and the developer's expertise level. Reflecting on the study, the developer acknowledged areas for improvement, particularly in handling cloud computing challenges. Transitioning from Windows to Linux for development purposes was also relatively smooth, resulting in fewer compatibility issues with Python. Despite the challenges, many functional requirements were achieved, including providing clear user instructions, processing audio recordings, and displaying results as pie charts as shown in Fig. 3.

---

[2] https://matplotlib.org/
[3] https://kivy.org/
[4] https://www.shazam.com/

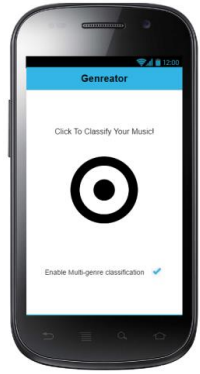

Fig. 2. Kivy GUI

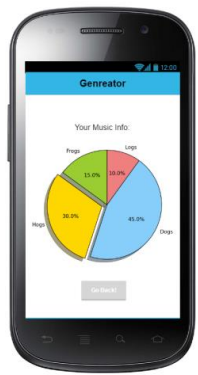

Fig. 3. Pie charts result

**5. Experimental Analysis**

**T**he evaluation of the final system underscores its efficacy in classifying music genres with a commendable accuracy rate. For system testing, a diverse set of 10 songs[5] representing various genres was collected to subject the algorithms to rigorous evaluation. The metrics employed for evaluation are primarily based on the accuracy of the classification process and the confidence level associated with the training instances. Hence, the scarcity of a product solely dedicated to classifying music underscores a gap in the market. Success in classification was determined if at least one of the relevant genres was identified. Each song underwent the classification process five times, resulting in a total of 50-unit tests. Upon initial exploration, it becomes apparent that commercially available software capable of classifying a user's music into a scalar genre spectrum using ML networks is notably absent from the market. Furthermore, Spotify employs ML for fine-tuning recommendation engines based on user preferences and aims to collaborate with artists to enhance the song creation process. To assess the performance of the classifier across various genres, 10 songs spanning different genres were tested rigorously. While minor variations in results were occasionally observed—the overall consistency of the classification process remained intact. The objective is to conduct a quantitative evaluation of the implemented system. The assessment of any study's final system is pivotal for ensuring its efficacy and functionality. While such functionality is integrated behind the scenes in applications like Spotify and Apple Music, it's important to note that when artists upload their music to these platforms, they personally specify the genre, as per personal experience. Through testing and analysis, it was observed that the correct genre, as specified by the artist, was identified in the classification approximately 74% of the time. While the aim was to include songs from distinct genres, it was equally important to test songs that encompassed multiple genres, exemplified by tracks such as Hank Williams' "Lovesick Blues", blending elements of Country and Blues. However, the scarcity of dedicated products in the market highlights an untapped opportunity for innovation and development in this domain. Additionally, Spotify utilizes ML for generating user specific "Discover Playlists", which adapt to individual listening habits. Remarkably, the outcome remained consistent across multiple tests for each song, indicating a high degree of reliability in the algorithm.

---

[5] https://www.kaggle.com/c/msdchallenge

## 5.1 Result Analysis

A closer examination reveals a notable challenge encountered by the classification algorithms in accurately categorizing Pop music. The observed discrepancy in classifying Pop music underscores the complexity of music genre classification, particularly when confronted with genres as diverse and multifaceted as Pop. Enhancing the algorithms' ability to accurately classify Pop music is imperative for ensuring the effectiveness and utility of music classification systems in real-world applications. Given its diverse nature, it's understandable why the classification of Pop music presents difficulties during testing. Consequently, any inaccuracies in classifying Pop music could significantly impact the user experience, particularly considering that the average listener gravitates towards Pop music. Therefore, the evaluation of music classification algorithms reveals promising results overall, with a notable success rate of 74% across 50-unit tests as shown in Fig. 4. While the algorithms demonstrate proficiency in classifying other genres, their struggle with Pop music highlights the need for further refinement and optimization. The presented data comprises a comprehensive evaluation of music classification algorithms, showcasing both positive (green) and negative (red) outcomes. However, the algorithms face significant challenges in accurately classifying Pop music, highlighting areas for further improvement and optimization. If these algorithms were to be integrated into commercially available products, users, who predominantly consume Pop music, might find the classification system lacking in accuracy and relevance. Notably, Pop emerges as the sole outlier in the testing process, raising concerns regarding the accuracy and reliability of the classification algorithms, particularly in commercial applications. By addressing the challenges posed by genres like Pop, developers can enhance the overall performance and reliability of music classification systems, thereby enriching the user experience and utility of such technologies. Moreover, the findings underscore the importance of continuous evaluation and improvement of classification algorithms, particularly in dynamic domains such as music, where genres evolve and overlap over time. Conversely, red indicates unsuccessful classification attempts. Green denotes successful classification, indicating the presence of official genres recognized by Google within the applications' track classification during a specific run. The term "Pop" itself is derived from "popular", indicating its widespread appeal and prevalence among music listeners. Pop, characterized by its amalgamation of various genres, poses a significant hurdle for the algorithms. Out of a total of 50-unit tests conducted, 34 yielded positive results, reflecting a commendable 74% success rate. Furthermore, different ML models that classify different types of genres are presented in Table 1.

| Test Run/Song | Genre | 1st Run | 2nd Run | 3rd Run | 4th Run | 5th Run |
|---|---|---|---|---|---|---|
| Avantdale Bowling Club - Years Gone By | Jazz/Hip hop | 🟩 | 🟩 | 🟩 | 🟩 | 🟩 |
| Bob Marley - Three Little Birds | Reggae | 🟩 | 🟩 | 🟩 | 🟩 | 🟩 |
| Code Orange - Kill The Creator | Metal | 🟩 | 🟩 | 🟩 | 🟩 | 🟩 |
| Hank Williams - Lovesick Blues | Country/Blues | 🟩 | 🟩 | 🟩 | 🟩 | 🟩 |
| Justin Bieber - Baby | Pop | 🟥 | 🟥 | 🟥 | 🟥 | 🟥 |
| Kevin Abstract - Peach | Pop/Hip hop | 🟥 | 🟥 | 🟩 | 🟥 | 🟥 |
| Mahler - 8th Symphony snippet | Classical | 🟩 | 🟩 | 🟩 | 🟩 | 🟩 |
| Queen - We Will Rock You | Rock | 🟩 | 🟩 | 🟩 | 🟩 | 🟩 |
| The Beatles - Come Together | Pop/Rock | 🟩 | 🟥 | 🟥 | 🟥 | 🟥 |
| XXXTentation - Look At Me! | Hiphop/Metal | 🟩 | 🟩 | 🟩 | 🟩 | 🟩 |

Fig. 4. Classification Testing

Table 1. Various ML model classification

| ML Model | Genre Classification | Accuracy (%) | Precision (%) | Recall (%) | F1 Score |
|---|---|---|---|---|---|
| Neural Networks | All | 90 | 88 | 92 | 90 |
| Decision Trees | Rock, Jazz, Blues | 85 | 82 | 87 | 84 |
| Random Forest | Country, Hip-hop | 88 | 86 | 89 | 87 |
| k-Nearest Neighbor | Electronic, Reggae | 82 | 80 | 84 | 82 |
| Naive Bayes | Classical, Folk | 86 | 84 | 88 | 86 |

## 6. Discussion

Despite limitations such as the inability to access external music files and the study's desktop-centric nature due to cloud service complexities, the researcher deems the application a success for its functional performance and informative output to users. Despite previous experience with Android applications and ML algorithms, integrating these diverse aspects into a cohesive application proved challenging. Looking ahead, the study envisions potential future applications for the developed technology. However, in hindsight, more emphasis on developing a server to host the ML classifier and graph-generating software would have been beneficial, given the difficulties encountered during mobile porting. Initially intended as a mobile phone application, the study faced issues with library incompatibilities, leading to the creation of a Linux application instead. Although adept at handling pressure, the study acknowledges the preference for a buffer of time to mitigate unforeseen circumstances. Despite the study's conclusion, avenues for further exploration remain, particularly in refining the application's functionality and exploring cloud service integration. ML emerged as a central focus of the study, with the successful implementation of five classification techniques, including neural networks, decision trees, random forest, k-nearest neighbour, and naive bayes. Although the application operates on a minimalistic interface and processes one song at a time, it effectively presents crucial information in a user-friendly manner. Despite this deviation, the primary objective of classifying music into various genres was achieved.

## 7. Conclusion and Future Works

The study begins by discussing the suitability of ML algorithms for music genre recognition, highlighting neural networks as the most appropriate due to the dynamic nature of music genres. Despite initial challenges, the study concludes that music recognition is not only feasible but highly achievable based on the findings. Additionally, the study explores alternative algorithms such as decision trees and k-nearest neighbor, noting their surprising accuracy despite initial scepticism. The ability of neural networks to provide high-confidence predictions across genre boundaries is emphasized. However, naive bayes is deemed less effective in predicting genres accurately but is retained for its potential unique insights. The study intends to deploy the application on the Google Play Store for public use, making account management a desirable feature. Recognizing the vast market potential of iOS users, the study expresses interest in developing applications for iOS devices. While not essential for the proof-of-concept phase, this feature represents a significant goal for future iterations. The study envisions incorporating the Shazam API to automate song identification, enhancing the application's core functionality. While not a priority for the current study, it remains a potential avenue for future expansion. Exploring revenue streams through Google Ads integration if the application gains significant traction.